\documentclass[preprint2]{aastex}
   
\usepackage{natbib}

\shorttitle{Oscillatory Reconnection}
\shortauthors{McLaughlin {\it{et al.}}}

\begin{document}

\title{Generation of quasi-periodic waves and flows in the solar atmosphere by oscillatory reconnection}

\author{J.~A. {McLaughlin}\altaffilmark{1} and G. {Verth}\altaffilmark{1}}
\affil{School of Computing, Engineering and Information Sciences, Northumbria University, Newcastle Upon Tyne, NE1 8ST, UK}

\author{ V. {Fedun}\altaffilmark{2} and  R. {{Erd{\'e}lyi}}\altaffilmark{2}}
\affil{Solar Physics and Space Plasma Research Centre (SP2RC), School of Mathematics and Statistics, University of Sheffield, Hounsfield Road, Hicks Building, Sheffield, S3 7RH, UK}

\email{james.a.mclaughlin@northumbria.ac.uk; gary.verth@northumbria.ac.uk; v.fedun@sheffield.ac.uk; robertus@sheffield.ac.uk}


\begin{abstract}

{We investigate the long-term evolution of an initially buoyant magnetic flux tube emerging into a gravitationally-stratified coronal hole environment and report on the resulting oscillations and outflows. We perform 2.5D nonlinear numerical simulations,  generalizing  the models of McLaughlin et al. (\citealp{McLaughlin2009}) and Murray et al. (\citealp{Murray2009}). We find that the physical mechanism of  {\emph{oscillatory reconnection}} naturally generates  quasi-periodic vertical outflows, with a transverse/swaying aspect. The vertical outflows consist of both a periodic aspect and  evidence of a positively-directed flow. The speed of the vertical outflow ($20-60\:$km/s) is comparable to those reported in the observational literature. We also perform a parametric study varying the magnetic strength of the buoyant flux tube and find a range of associated periodicities:  $1.75-3.5\:$min. Thus, the mechanism of oscillatory reconnection may provide a physical explanation to some of the high-speed, quasi-periodic, transverse outflows/jets recently reported by a multitude of authors and instruments.}

\end{abstract}



\section{Introduction}\label{sec:Introduction}

Improvements in the spatial and temporal resolution of solar observations  have led to a recent  deluge in reported magnetohydrodynamic (MHD)  wave motions (e.g. see Nakariakov \& Verwichte \citealp{Nakariakov2005}; De Moortel \citealp{Ineke2005};    Banerjee et al. \citealp{Banerjee2007};   Ruderman \& Erd\'elyi \citealp{Misha2009};    Goossens et al. \citealp{Goossens2011};          McLaughlin et al. \citealp{McLaughlinREVIEW} for a recent list). Here, we focus on observations of transverse motions in the solar atmosphere (e.g. Tomczyk et al. \citealp{Tomczyk2007}; De Pontieu et al. \citealp{Bart2007}; Cirtain et al. \citealp{Cirtain2007}; Erd\'elyi \& Taroyan \citealp{Robertus2008}; He et al. \citealp{He2009a}; \citealp{He2009b}; Liu et al. \citealp{Liu2009}; \citealp{Liu2011};  Morton et al. \citealp{Morton2011}; Okamoto et al. \citealp{Okamoto2011}). These transverse motions have been called Alfv\'en waves by some authors,  although this is subject to debate and they are  alternatively interpreted as kink waves (e.g. see arguments by  Erd\'elyi \& Fedun \citealp{RobertusViktor}; Van Doorsselaere et al. \citealp{Van2008}). The dispute rests not with the observations themselves, but with the appropriate interpretation: MHD wave modes of an overdense cylinder versus MHD waves of a homogeneous plasma. These arguments and others, e.g.  whether or not a stable waveguide actually exists in the solar atmosphere, are not the focus of this current paper.

Tomczyk et al. (\citealp{Tomczyk2007}) utilised the {\emph{CoMP/Coronal Multi-channel Polarimeter}} instrument to report on ubiquitous, small-amplitude, transverse disturbances, propagating along magnetic field lines. The authors do not report on how the waves are generated, but do speculate the waves may originate from within the chromospheric network that forms the footpoints of the observed loops. De Pontieu et al. (\citealp{Bart2007}) used {\emph{Hinode/SOT}} measurements in an attempt  to reveal Alfv\'en/transversal  waves in the chromosphere with strong amplitudes ($10-30\:$km/s) and periods $100-500\:$seconds. Ca II H-line images also reveal a plethora of dynamic, jet-like extrusions called chromosphere spicules, or {\emph{type II spicules}}. These spicules undergo a swaying/oscillatory motion perpendicular to their own axis, which the authors described as Alfv\'enic motions. Again, this interpretation is disputed by other authors (e.g. He et al. {\citealp{He2009b}}; Verth et al. {\citealp{Verth2011}}) who interpreted these spicule oscillations as kink waves,  due to the fact that spicules are overdense in comparison with the ambient plasma.


De Pontieu et al. (\citealp{Bart2011}) report on a link between chromospheric spicules/jets and their coronal spicules/jets counterparts, i.e. suggesting a mechanism for imparting chromospheric plasma into the corona. The coronal spicules are strongly heated and are seen to rapidly propagate upwards, but the authors report that there are currently no models for what drives and heats the observed jets (see also a review by Sterling \citealp{Sterling2000}). Okamoto \& De Pontieu (\citealp{Okamoto2011}) report on the statistical properties of transverse (Alfv\'enic)  waves along spicules and report median velocity amplitudes and periods  of $7.4\:$km/s and $45\:$seconds, respectively (see also review by Zaqarashvili \& Erd\'elyi \citealp{Zaq2009}). McIntosh et al. (\citealp{McIntosh2011}) reported transition region observations of ubiquitous, transverse (swaying/Alfv\'enic) motions that are outwardly propagating, with amplitudes $\sim20\:$km/s and periods $\sim 100-500\:$seconds,  energetic enough to heat the fast solar wind. Again, the authors note that the challenge remains to understand how and where these waves are generated in the solar atmosphere.

Thus, transverse/swaying motions have been observed over a range of temperatures, wavelengths, speeds and scales. However, the origin of these propagating, transverse oscillations remains a mystery. Liu et al.  (\citealp{Liu2011}) summarise possible generation mechanisms including an oscillating wake from a CME or periodic reconnection (e.g. Chen \& Priest \citealp{Chen2006}; Sych et al. \citealp{Sych2009}).

\subsection{Waves versus Flows Interpretation}\label{sec:flows}

In addition to transverse MHD wave observations, slow MHD waves have also been observed in the solar atmosphere (e.g. Ofman et al. \citealp{Ofman1997}; DeForest \& Gurman \citealp{DeForest1998}; Berghmans \& Clette \citealp{Berghmans1999}; De Moortel et al. \citealp{Ineke2000}; Ofman \& Wang \citealp{Ofman2008}; Erd\'elyi \& Taroyan \citealp{Robertus2008}). More recently, this propagating slow wave interpretation has been challenged. De Pontieu \& McIntosh (\citealp{Bart2010})  show that {\emph{Hinode/EIS}} measurements of intensity and velocity oscillations of coronal lines are driven by a quasi-periodically varying component in the blue wing of the emission line, i.e. providing evidence of  quasi-periodic upflows. Moreover, such upflows ($\sim 50-150\:$km/s in the line-of-sight) have also been reported in coronal loop footpoints in active regions (De Pontieu et al. \citealp{Bart2009c}) and quiet-Sun regions (McIntosh \& De Pontieu \citealp{McIntosh2009a}). These authors also suggest a direct link between these outflows/propagating disturbances and chromospheric spicules, i.e. the fountain-like jets that protrude into the corona (McIntosh \& De Pontieu \citealp{McIntosh2009b}).

McIntosh et al. (\citealp{McIntosh2010}) analysed {\emph{STEREO}} observations of polar plumes and identified high-speed upflows, again as opposed to previous interpretations as propagating slow waves. These authors observed high-speed jets travelling along the plumes with  mean velocity of $135\:$km/s and  repeating quasi-periodically (repeat times 5-25 minutes). Using {\emph{SDO/AIA}} to extend this analysis  to plume-like structures originating from equatorial coronal holes and quiet-Sun regions, Tian et al. (\citealp{Tian2011}) found that the outflows are not restricted to plumes. These authors reported that the outflows exhibit transverse, swaying motions and probably  originate from the magnetic network of the quiet-Sun and coronal holes. In contrast, Verwichte et al. (\citealp{Verwichte2010}) argue that a slow mode interpretation remains a valid explanation for the observed quasi-periodic intensity perturbations, and show that slow waves inherently have a bias towards emission in the blue wing of the emission line due to the in-phase nature of the velocity and density perturbations.

Thus, it is clear that there is a need for a better understanding of the physical generation mechanism for these ubiquitous,   transverse, propagating motions, and that both quasi-periodic motions and possible flows must be kept in mind. This paper attempts to describe such a physical mechanism by utilising a dynamic,  reconnection numerical model.

\subsection{Flux Emergence \& Oscillatory Reconnection}\label{sec:flux_emergence_INTRO}

Magnetic field is continuously emerging on the Sun over a range of scales (see Archontis \citealp{ArchontisREVIEW} for a comprehensive review). Magnetic flux tubes, formed at the tachocline, rise buoyantly through the convection zone. As they reach the photosphere, their buoyant-rise ends and an instability allows the magnetic flux to penetrate into the solar atmosphere. The subsequent evolution of the newly-emerged flux is then dominated both by the properties of the rising flux tube itself and the pre-existing magnetic topology the tube emerges into. Flux emergence is well described in the existing literature, as is the dynamic reconnection associated with the collision of newly-emerged flux and pre-existing magnetic field (e.g. Shibata et al. \citealp{Shibata1992}; Yokayoma \& Shibata \citealp{Yokoyama1995}; \citealp{Yokoyama1996}; Archontis et al. \citealp{Vasilis2004}; \citealp{Vasilis2005}; \citealp{Vasilis2006}; \citealp{Vasilis2007}; Isobe et al. \citealp{Isobe2005}; \citealp{Isobe2006}; Murray et al. \citealp{Murray2006}; {{Galsgaard et al.}} \citealp{Galsgaard2007}; Moreno-Inserti et al. \citealp{Moreno2008}, and references therein).

Shibata et al. (\citealp{Shibata2007}) reported {\emph{Hinode/SOT}} observations of the ubiquitous presence of chromospheric anemone jets (velocity $10-20\:$km/s). These numerous, small-scale jets, seen in, e.g., Ca II H broadband, display an inverted Y-shape, i.e. the characteristic shape of anemone jets (e.g. Shibata et al. \citealp{Shibata1994}; Yokoyama \& Shibata \citealp{Yokoyama1995}).  The anemone shape is formed as a result of magnetic reconnection between an emerging magnetic bipole and a pre-existing vertical field.

Reconnection can occur when strong currents cause the magnetic fieldlines to diffuse through the plasma and change the connectivity (Parker \citealp{Parker1957}; Sweet \citealp{Sweet1958}; Petsheck \citealp{Petschek1964}). In 2D, reconnection can only occur at null points (Priest \& Forbes \citealp{Priest2000}). Dungey (\citealp{Dungey1953}) reported that a perturbed X-point can collapse if the footpoints are free to move,  Mellor et al. (\citealp{Mellor2002}) studied the linear collapse of a 2D null point, and Imshennik \& Syrovatsky (\citealp{Imshennik1967}) described the collapse with an exact, nonlinear solution of the ideal MHD equations. However, these papers do not include the effect of gas pressure, which acts to limit the growth of the current density. In considering the relaxation of  a 2D X-type null point, Craig \& McClymont (\citealp{Craig1991}) found that free magnetic energy is dissipated by the phenomenon of {\emph{oscillatory reconnection}}, which couples resistive diffusion at the null to global advection of the outer field. McLaughlin et al. (\citealp{McLaughlin2009}) investigated the behavior of nonlinear fast magnetoacoustic waves near a 2D X-point and found that the incoming wave deforms the null point into cusp-like point which in turn collapses to a current sheet. The system then evolves periodically through a series of horizontal/vertical current sheets  with associated changes in connectivity, i.e. the system displays {{oscillatory reconnection}}. Longcope \& Priest (\citealp{Longcope2007}) investigated the diffusion at the null of a 2D current sheet subjected to a suddenly enhanced resistivity, finding that the diffusion couples to a fast mode which propagates the current away at the local Alfv\'en speed.

Of particular importance to the work presented in this paper is that of Murray et al. ({\citealp{Murray2009}}). Flux emerging into a pre-existing field has been studied in great detail before, but Murray et al.  ({\citealp{Murray2009}}) were the first to investigate the long-term evolution of such a  system, i.e. many previous simulations end once reconnection is first initiated. Murray et al. utilised a stratified atmosphere permeated by a unipolar magnetic field (representing a coronal hole) and investigated the emergence of a  buoyant flux tube. Murray et al. found that a series of \lq{reconnection reversals}\rq{ }  take place as the system searches for equilibrium, i.e. a cycle of inflow/outflow bursts followed by outflow/inflow bursts. Thus, the system demonstrates {oscillatory reconnection} (e.g. Craig \& McClymont  \citealp{Craig1991};  McLaughlin et al. \citealp{McLaughlin2009}), initiated in a self-consistent manner.  Murray et al. also detail the physics behind the phenomena.

\begin{figure*}
\epsscale{1.7}
\plotone{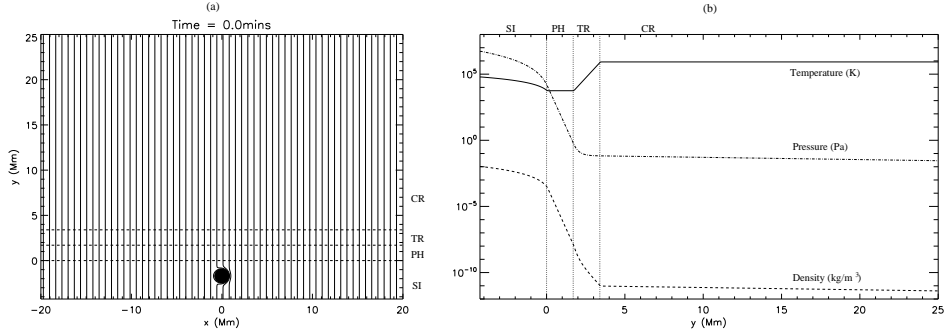}
\caption{$(a)$ Equilibrium magnetic field: vertical magnetic field representing a unipolar coronal hole and curved fieldlines of our initial flux tube, centered at $(x,y)=(0,-4.25)\:$Mm. $(b)$ Equilibrium conditions of the numerical model: temperature (solid), gas pressure (dashed), density (dot-dashed). In both figures, the solar interior (SI), photosphere (PH), transition region (TR) and corona (CR) are indicated by the dotted lines. Only a subset of the full numerical domain is shown.}
\label{Figure1}
\end{figure*}

The aim of this paper is to further generalize the model of Murray et al.  ({\citealp{Murray2009}}) and to detail the oscillatory outputs due to oscillatory reconnection (Craig \& McClymont  \citealp{Craig1991};  McLaughlin et al. \citealp{McLaughlin2009}). We will also investigate the dependency and robustness of the model by varying the initial magnetic strength of the buoyant flux tube.

The paper has the following outline: the numerical model is detailed in \S\ref{sec:model}, brief recall of Murray et al.  ({\citealp{Murray2009}}) is described in \S\ref{sec:fluxemergence} and the quasi-periodic outputs are reported in \S\ref{sec:outflows}. \S\ref{sec:parameter_study} investigates the dependency of  the model to the initial strength of the buoyant flux tube and conclusions are presented in \S\ref{sec:conclusions}.


\section{Numerical Model}\label{sec:model}

We consider the two-dimensional, nonlinear, compressible, resistive MHD equations, including gravitational effects:
\begin{eqnarray}
 \rho \left[ {\partial {\bf{v}}\over \partial t} + \left( {\bf{v}}\cdot\nabla \right) {\bf{v}} \right] &=&- \nabla p + \left( {{\frac{1}{\mu}}}   \nabla \times {\bf{B}}  \right)\times {\bf{B}}   +\rho {\bf{g}}   \; ,\nonumber \\
 {\partial {\bf{B}}\over \partial t}  &=& \nabla \times \left ({\bf{v}}\times {\bf{B}}\right ) + \eta \nabla ^2  {\bf{B}}\; ,\nonumber \\
 \rho \left[{\partial {\epsilon}\over \partial t}  + \left( {\bf{v}}\cdot\nabla \right) {\epsilon}\right] &=& - p \nabla \cdot {\bf{v}} + {{\frac{1}{\sigma}}} \left| {\bf{j}} \right| ^2 + Q_{\rm{shock}} \; \nonumber  ,\\
{\partial \rho\over \partial t} + \nabla \cdot \left (\rho {\bf{v}}\right ) &=& 0\; , \label{MHDequations}  
\end{eqnarray}

where $\rho$ is the mass density, ${\bf{v}}$ is the plasma velocity, ${\bf{B}}$ the magnetic induction (usually called the magnetic field), $p$ is the plasma pressure,  $ \mu = 4 \pi \times 10^{-7} \/\mathrm{Hm^{-1}}$  is the magnetic permeability, acceleration due to gravity ${\bf{g}} = -g {{\hat{\bf{y}}}}$,  $\sigma$ is the electrical conductivity,  $\eta=1/ {\mu \sigma} $ is the magnetic diffusivity, $\epsilon= {p / \rho \left( \gamma -1 \right)}$ is the specific internal energy density,   $\gamma={5 / 3}$ is the ratio of specific heats and ${\bf{j}} = {{\nabla \times {\bf{B}}} / \mu}$ is the electric current density. 

We solve these governing equations using the {\emph{LARE2D}} numerical code (Arber et al. \citealp{Arber2001}) which utilizes artificial shock viscosity to introduce dissipation at steep gradients, and the details of this technique, often called Wilkins  viscosity, can be found in Wilkins (\citealp{Wilkins1980}). Thus, $Q_{\rm{shock}}$ represents the viscous heating at shocks. Heat conduction and radiative effects are neglected in the present study.

\begin{figure*}[t]
\epsscale{2.5}
\hspace{-4.0cm}
\plotone{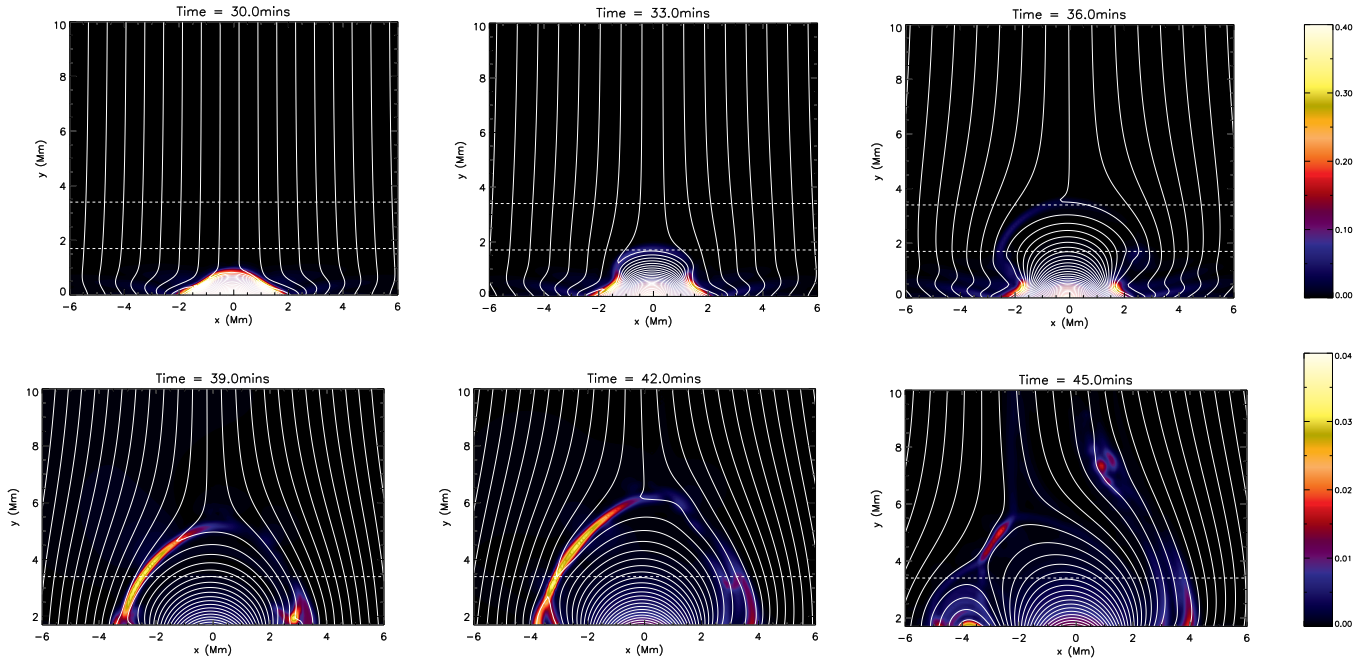}
\hspace{-4.0cm}
\caption{Contours of current density (${|}{\bf{J}}{|}$, units A$\:$m$^{-2}$) and selection of fieldlines at times $t=30$, $33$, $36$, $39$, $42$ and  $45\:$mins. Dotted lines denote horizontal layers of the equilibrium solar atmosphere. Note that the numerical domain plotted and colour bars change between the two rows.}
\label{Figure2}
\end{figure*}


We now introduce a change of scale to non-dimensionalise all variables. Letting ${\rm{\bf{v}}} = {\rm{v}}_0 {\mathbf{v}}^*$,  ${\mathbf{B}} = B {\mathbf{B}}^*$, $x = L x^*$, $y=L y^*$, $\rho={\rho}_0 \rho^*$, $p = p_0 p^*$, $\nabla = \nabla^* / L$, $t={t}_0 t^*$,  ${\bf{A}}=B L {\bf{A}}^*$, $g = g_0$  and $\eta = \eta_0$, where * denotes a dimensionless quantity and ${\rm{v}}_0$, $B$, $L$, ${\rho}_0$, $p_0$, ${t}_0$, $g_0$ and $\eta_0$ are constants with the dimensions of the variable they are scaling. Here, ${\bf{A}}=A_z\: {\hat{\bf{z}}}$ is the vector potential. We then set $ {B} / {\sqrt{\mu \rho _0 } } ={\rm{v}}_0$ and ${\rm{v}}_0 =  {L} / {{t_0}}$ (this sets ${\rm{v}}_0$ as a constant background Alfv\'{e}n speed). We also set ${\eta_0 {t}_0 } /  {L^2} =R_m^{-1}$, where $R_m$ is the magnetic Reynolds number. This process non-dimensionalises equations (\ref{MHDequations}) and under these scalings, $t^*=1$ (for example) refers to $t={t}_0=  {L} / {{\rm{v}}_0}$; i.e. the time taken to travel a distance $L$ at the background Alfv\'en speed. For the rest of this paper, we drop the star indices; the fact that all variables are now non-dimensionalised is understood.

The values returned from  equations (\ref{MHDequations}) are made dimensional using the following choice of solar constants:  (photospheric) pressure scale height $L=1.7\times 10^5\:$m, time $t_0=25\:$seconds, velocity ${\rm{v}}_0={L} / {{t_0}}=6.8\times 10^3\:$m/s, density ${\rho}_0=3.0\times 10^{-4}\:$kg/m$^3$, pressure  $p_0=1.2\times 10^4\:$Pa, temperature $T_0=5.6\times 10^3\:$K, magnetic field $B=1.3\times 10^3\:$G and gravity $g_0=270\:$m/s.


\subsection{Initial and Boundary Conditions}\label{sec:setup}

To simulate flux emerging into the solar atmosphere, we follow the numerical set-up of Murray et al. (\citealp{Murray2009}). We consider a numerical domain comprising of four horizontal layers (Figure \ref{Figure1}a). Above the lower boundary, we have a solar interior that is marginally stable to convection, a $5600\:\:$K isothermal photosphere, a transition region with a rapid (power law) temperature increase  and, above this, an isothermal corona. Each of the layers is initially in hydrostatic equilibrium. Figure \ref{Figure1}b shows the equilibrium conditions.


To model a unipolar corona hole, we chose an initial magnetic field to be a negatively-directed vertical magnetic field of $19.5\:$G. The coronal hole temperature, density and field strength are taken from Zhang et al. ({\citealp{Zhang2007}}) and Baker et al. (\citealp{Baker2009}).

To initiate flux emergence, a magnetic flux tube is placed in the solar interior at $x=0$ and a depth of $y=-1.7\:$Mm. In cylindrical coordinates, the magnetic flux tube is chosen to have ${\bf{B}}=B_0\left(0, \alpha r e^{-r^2/R^2}, e^{-r^2/R^2}\right)$. We choose $B_0=3.25\times 10^3\:$G, $R=4.25\times 10^5\:$m, and {{$\alpha=-0.064\times 2\pi$}} for each $L$ length in the axial direction. The buried flux tube is set in radial force balance and in thermal equilibrium with the external plasma. Thus, a density difference exists such that the flux tube is buoyant relative to its surroundings and, at the start of the simulation, will rise bodily upwards/towards the photosphere.

We chose a numerical domain $-68\:$$\:\le x\le 68\:\:$Mm, $-4.25\:$$\:\le y \le 63.75\:\:$Mm  using $1600\times 832$ gridpoints of uniform spacing. All boundaries are fixed in $x$ and $y$. Convergence testing was carried out with double and half the resolution.


\section{Flux Emergence \& Recreation of Murray et al. results}\label{sec:fluxemergence}

\begin{figure*}
\epsscale{2.5}
\hspace{-4.0cm}
\plotone{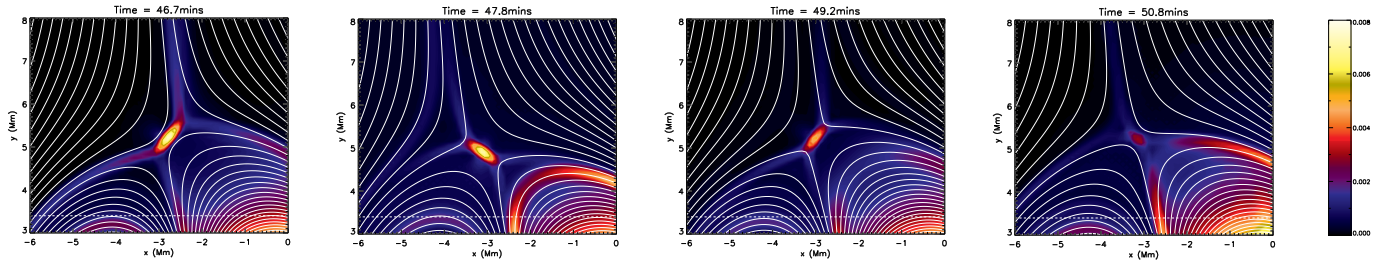}
\hspace{-4.0cm}
\caption{Contours of current density (${|}{\bf{J}}{|}$, units A$\:$m$^{-2}$) and selection of fieldlines at times $t=46.7$, $47.8$, $49.2$ and $50.8\:$mins. Dotted lines denote a horizontal layer of our equilibrium solar atmosphere (i.e. change from transition region to coronal temperature profile). Note that we plot different axes compared to Figure \ref{Figure2}.}
\label{Figure3}
\end{figure*}

Flux emergence is well documented in existing literature (e.g. Murray et al. \citealp{Murray2006}, Archontis et al. \citealp{ArchontisREVIEW}, as well as the papers listed in \S$\ref{sec:flux_emergence_INTRO}$ and references therein). Figure \ref{Figure2} (top row) shows that  the buoyant magnetic tube rises and emerges into the model atmosphere,  and that the emerging flux compresses the pre-existing magnetic field as it expands.  To the north-west side of the emerging flux, the magnetic field is directed positively out of the solar surface whereas the neighboring coronal hole is directed in the opposite direction. Thus, a current sheet builds up at this interface between the two flux systems and this can be clearly seen in the bottom row of Figure \ref{Figure2}. Reconnection commences at  $t=41.6\:$mins.



The system demonstrates the phenomenon of  {\emph{oscillatory reconnection}} as it searches for an equilibrium. In Figure \ref{Figure3}, we see that at $t=46.7\:$mins the system forms a current sheet located around $-3.5<x<-2.5\:$Mm, $4.5<y<-5.5\:$Mm, at an angle of approximately $\pi/4$ relative to the positive $x-$direction. In this paper, we shall refer to a current sheet at this angle as an {\emph{orientation 1}} current sheet. At $t=47.8\:$mins, we see that a new current sheet has formed at a similar location, but now at an angle of approximately $3 \pi/4$ relative to the positive $x$-direction. We shall refer to a current sheet formed at this  angle   as an {\emph{orientation 2}} current sheet. At  $t=49.2\:$mins,  we see that a current sheet has formed again in a similar location, but that this current sheet is again of orientation 1. Finally, at  $t=50.8\:$mins, we see the formation of an     orientation 2 current sheet, again in a similar location (it is also clear that this current sheet is weaker, i.e. ${|}{\bf{J}}{|}_{\rm{max}}$ is decreased,  compared to the previous current sheets). Thus,  Figure \ref{Figure3} illustrates the formation of a cycle of current sheets, i.e. the formation of orientation 1, followed by orientation 2, followed by orientation 1 again, and so on. Note that this figure is a qualitative illustration of the periodic nature of the current sheet formation in this system (\S\ref{sec:period}  will provide quantitative evidence). Note that our terminology of {\emph{orientation 1}} and {\emph{2}} is purely arbitrary;  a similar periodic cycle of current formation was seen in McLaughlin et al. (\citealp{McLaughlin2009}) and was referred to a cycle of {\emph{horizontal}} and {\emph{vertical}} current sheets.

Thus, we recover the results of Murray et al. (\citealp{Murray2009}).  Murray et al. demonstrated that, using fieldline tracing, one  can quantitatively demonstrate the periodic change in connectivity of the open flux over time, i.e. evidence of reconnection. We recover the fieldline-tracing results of Murray et al.  (their Fig. 4). Hence, we  use the terminology {\emph{oscillatory reconnection}} to refer to the periodic formation of orientation 1 and 2 current sheets as well as the associated periodic changes in connectivity. We now focus on the observable consequences and outputs of such an evolving system.


\subsection{Generation of quasi-periodic outflows}\label{sec:outflows}

\begin{figure*}
\epsscale{2.6}
\hspace{-4.0cm}
\plotone{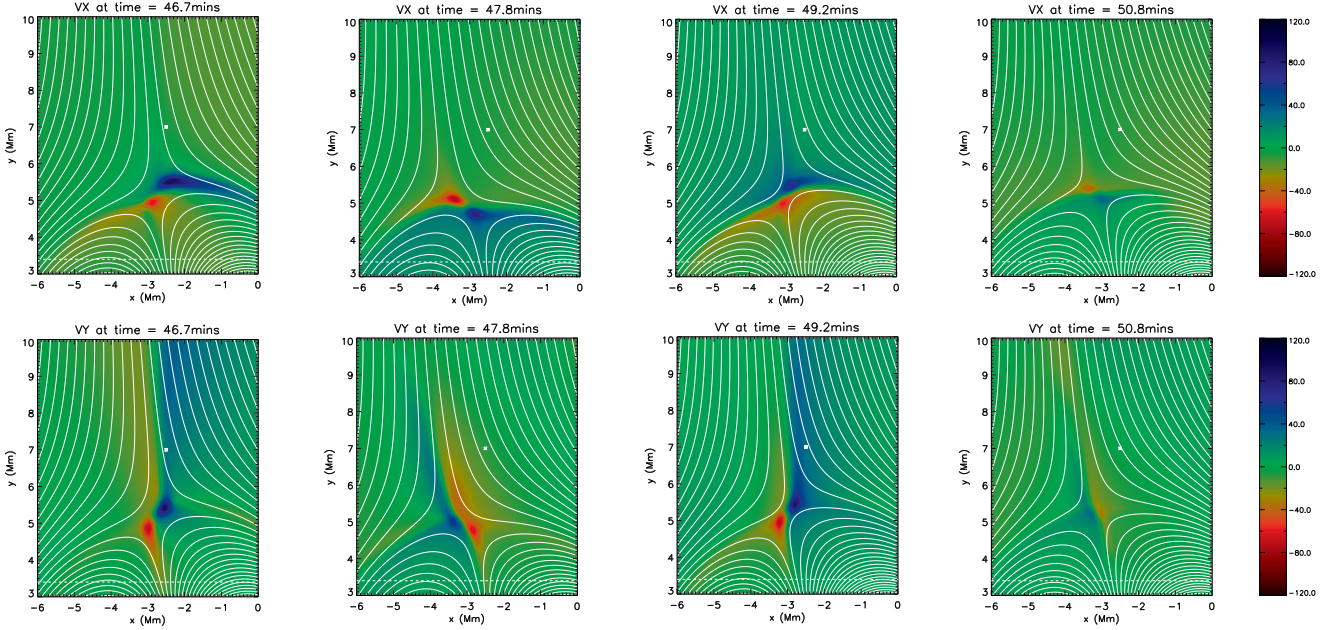}
\hspace{-4.0cm}
\caption{Contours of (top row) ${{{{v}}}}_x$ and (bottom row) ${{{{v}}}}_y$, in units of km/s, and selection of fieldlines at times $t=46.7$, $47.8$, $49.2$ and $50.8\:$mins. Blue/red corresponds to positive/negative motions. White dot indicates the fixed point $(x_0, y_0)=(-2.5\:$Mm, $7\:$Mm). Dotted line denotes a change from equilibrium transition region to coronal temperature profile.}
\label{Figure4}
\end{figure*}

As current sheets form and reconnection commences in the simulation, we observe strong outflows emanating from the ends of the current sheet, in agreement with classic steady-state reconnection theory (e.g. Sweet \citealp{Sweet1958}; Parker \citealp{Parker1957}; Petschek \citealp{Petschek1964}). Upon leaving the current sheet, these jets (strong outflows) collide with the magnetic field already in the outflow regions and are deflected into two secondary jets at angles of approximately $\pm \pi/4$ to the original jet  (a termination shock is also present). The schematic structure of these reconnection jets  is in good agreement with the description of Forbes (\citealp{Forbes1988}). Since these secondary jets are deflected at angles  of approximately $\pm \pi/4$ and  due to the orientation of the current sheets, they manifest themselves as either primarily horizontal and/or vertical outflow jets. {{For orientation 1 current sheets, jets from the lower end of the current sheet give rise (periodically) to negatively-directed ${{{{v}}}}_x$ and ${{{{v}}}}_y$ motion. Meanwhile, jets from the upper end of the current sheet give rise (periodically) to positively-directed ${{{{v}}}}_x$ and ${{{{v}}}}_y$ motion. For orientation 2 current sheets, the two jets have mixed velocity components, and this can be clearly seen in Figure \ref{Figure4}. Figure \ref{Figure4} shows}}  contours of ${{{{v}}}}_x$ (top row) and  ${{{{v}}}}_y$ (bottom row)  at four snapshots in our simulation (the blue/red colour table corresponds to positive/negative motions).


Let us first consider the  ${{{{v}}}}_x$ behaviour (top row of Figure \ref{Figure4}). The characteristic behaviour can be summarised as follows:
\begin{list}{$\bullet$}{\leftmargin=1em \itemindent=0em}
\item{At $t=46.7\:$mins  (orientation 1 current sheet), we have strong horizontal outflows, with positively-directed ${{{{v}}}}_x$ motions (blue) ejected from the upper end of the current sheet.}
\item{At $t=47.8\:$mins  (orientation 2 current sheet), we have  negatively-directed ${{{{v}}}}_x$ motions ejected  from the upper end of the current sheet.}
\item{This cycle repeats, and we have   positively-directed $v_x$ motions again at $t=49.2\:$mins  (orientation 1) followed by negatively-directed ${{{{v}}}}_x$ motions at $t=50.8\:$mins  (orientation 2).}
\end{list}
A similar pattern is observed emanating from the lower end of the current sheet but with the opposite orientation, i.e.  at $t=46.7\:$mins  (orientation 1) we have  negatively-directed ${{{{v}}}}_x$ motions, followed by  positively-directed ${{{{v}}}}_x$ motions at $t=47.8\:$mins (orientation 2), again in a repeating cycle.

\begin{figure*}
\epsscale{2.1}
\plotone{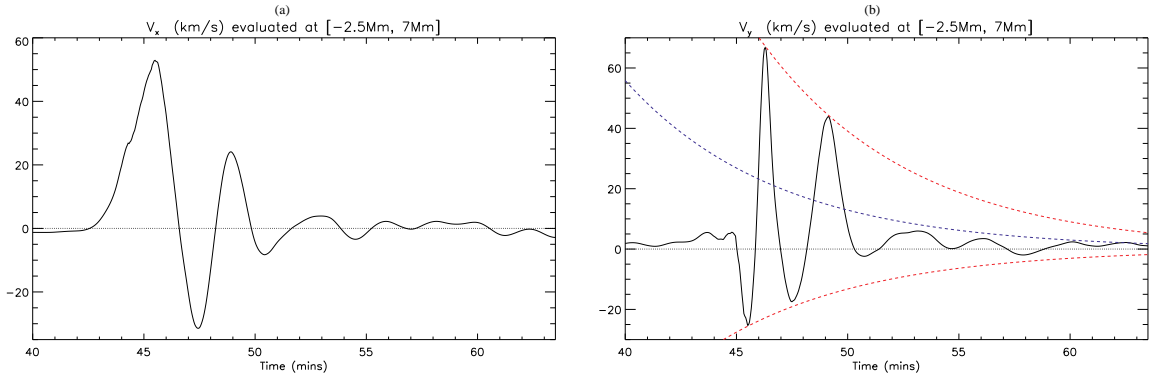}
\caption{$(a)$ Time evolution of ${{{{v}}}}_x$ (km/s) at the point $(x_0, y_0)=(-2.5\:$Mm, $7\:$Mm). $(b)$ Time evolution of ${{{{v}}}}_y$ (km/s) at  $(x_0, y_0)=(-2.5\:$Mm, $7\:$Mm). The red dashed lines indicate an exponentially damped envelope $\sim {{{{v}}}}_{y-{\rm{max}}} e^{-\lambda t}$ and  $\sim {{{{v}}}}_{y-{\rm{min}}} e^{-\lambda t}$ where $\lambda=0.1464\:{\rm{s}}^{-1}$. In both figures, the dotted line denotes zero velocity.}
\label{Figure5}
\end{figure*}


Let us now consider the  ${{{{v}}}}_y$ behaviour (bottom row of Figure \ref{Figure4}). The characteristic behaviour can be summarised as follows:
\begin{list}{$\bullet$}{\leftmargin=1em \itemindent=0em}
\item{At $t=46.7\:$mins  (orientation 1 current sheet), we have strong vertical outflows, with positively-directed ${{{{v}}}}_y$ motions (blue) ejected from the upper end of the current sheet. Interestingly, we also have negatively-directed (but weaker)  ${{{{v}}}}_y$ motion adjacent to (to the left of) our positively-directed ${{{{v}}}}_y$ outflow. This negatively-directed ${{{{v}}}}_y$ motion is associated with the inflow region of the simulated  current sheet, and thus acts at right-angles to the current sheet orientation.}
\item{At $t=47.8\:$mins  (orientation 2), we now have a positively-directed ${{{{v}}}}_y$ motions ejected from the upper end of the current sheet. Again,  negatively-directed ${{{{v}}}}_y$ motion appears adjacent to (now to the right of) our positively-directed ${{{{v}}}}_y$ outflow  (again  associated with  inflow into our reconnection region).}
\item{The cycle then repeats:  we have   positively-directed ${{{{v}}}}_y$ motions again at $t=49.2\:$mins  (orientation 1) and negatively-directed ${{{{v}}}}_x$ motions at $t=50.8\:$mins  (orientation 2).}
\end{list}
The vertical ${{{{v}}}}_y$ motions are of obvious interest for comparison with observations.  Following the terminology of  Murray et al. (\citealp{Murray2009}) we refer to these vertical ${{{{v}}}}_y$ motions as a {\emph{collimated jet}}.


\subsubsection{Transverse/swaying collimated jets}\label{sec:vertical_VX}

We now consider  these  vertical outflow jets and swaying, transverse motions in further detail. To do so, we measure the  ${{{{v}}}}_x$ and ${{{{v}}}}_y$ signal at a fixed point $(x_0, y_0)=(-2.5\:$Mm, $7\:$Mm) where this particular point has been chosen as it is located close to the upper end of the orientation 1 current sheet (slightly above and to the right) and thus is well placed to measure the transverse motions and ${{{{v}}}}_y$ outflows of the resultant jets.

By considering the evolution of the collimated  jet, we note that the central axis will be horizontally-displaced periodically as the current sheet contracts and lengthens (i.e. evolves between orientation 1 and 2, and back again). Whilst evolving from orientation 1 to 2, the collimated jet will appear to move in the negative $x-$direction (i.e. an orientation 1 current sheet first contracts, then lengthens into an orientation 2 current sheet). Conversely, whilst evolving from an orientation 2 current sheet (back) to orientation 1, the collimated jet will appear to move in the positive $x-$direction. This displacement will repeat itself as the cycle repeats. Thus, the collimated jet displays a characteristic swaying or transverse motion. Note that this transverse behaviour is specifically due to the oscillatory reconnection mechanism, and would be absent for a single, steady-state reconnection jet.

Qualitatively, this transverse displacement can be seen by comparing the locations of the (blue) collimated jet in Figure \ref{Figure4} (bottom row). However, such  displacement can also be quantitatively measured from our simulation. In Figure \ref{Figure5}a, we see the evolution of ${{{{v}}}}_x$ (km/s) at the fixed point $(x_0, y_0)=(-2.5\:$Mm, $7\:$Mm).  The oscillatory behaviour of ${{{{v}}}}_x$ can be clearly seen, i.e.  the quasi-periodic transverse/swaying displacement. Note that the change from orientation 1 to 2 occurs at $t \approx 47\:$mins. Before this time, the strong positive ${{{{v}}}}_x$ motion is associated with the initial formation of the orientation 1 current sheet.


\subsubsection{Quasi-periodic vertical outflows}\label{sec:vertical_VY}

Figure \ref{Figure5}b shows the evolution of ${{{{v}}}}_y$, i.e.  the vertical outflow,  at the fixed point  $(x_0, y_0)=(-2.5\:$Mm, $7\:$Mm). We can clearly see the oscillatory behaviour:  the outflow changes from ${{{{v}}}}_y<0$ at $t \approx 45.5\:$mins, to ${{{{v}}}}_y>0$ at $t \approx 46.5\:$mins, to ${{{{v}}}}_y<0$ at $t \approx 47.5\:$mins, to ${{{{v}}}}_y<0$  at $t \approx 49\:$mins. After $t \approx 50\:$mins, the signal is much weaker since the bulk of the (oscillatory) reconnection has occurred by this time.

We fit the oscillation with an  exponentially-damped envelope $\sim {{{{v}}}}_{y-{\rm{max}}} e^{-\lambda t}$ for ${{{{v}}}}_y>0$ and   $\sim {{{{v}}}}_{y-{\rm{min}}} e^{-\lambda t}$ for ${{{{v}}}}_y<0$ (red dashed lines), where  $\lambda=0.1464\:{\rm{s}}^{-1}$ is determined experimentally. It is also important to note that there is a stronger positive ${{{{v}}}}_y$ signal than negative ${{{{v}}}}_y$ signal, i.e. there is more \lq{upflow}\rq{} than \lq{downflow}\rq{}. This can be most clearly seen by calculating the average of the two exponentially decaying envelopes, i.e. $\sim 0.5 \left( {{{{v}}}}_{y-{\rm{max}}} + {{{{v}}}}_{y-{\rm{min}}}\right) e^{-\lambda t}$, and this average is denoted by the blue dashed line. Since this curve is positive, it indicates the presence of a positive upflow, i.e. for zero flow, we would expect the two exponentially decaying envelopes to average to zero.

{{Note that the generation of this  positively-directed vertical net flow is tightly localised to the upper ends of the (evolving) current sheet, and thus one must be careful when analysing the oscillatory signal at a fixed point (as in Figure \ref{Figure5}b). Similar oscillatory results and upflows are obtained for other choices of fixed points located around  $(x_0, y_0)=(-2.5\:$Mm, $7\:$Mm) and within the same (north-easterly) domain of connectivity.}}


We also note that for both Figures \ref{Figure5}a and  \ref{Figure5}b, the conditions before  $t\approx 43\:$mins are explained by the initial flux emergence itself, i.e. the flux tube is expanding upwards  and sideways  into the solar atmosphere and so, for $x_0 <0$, a small ${{{{v}}}}_y > 0$ and ${{{{v}}}}_x <0$ is initially present.

\begin{figure}[h]
\epsscale{1.0}
\plotone{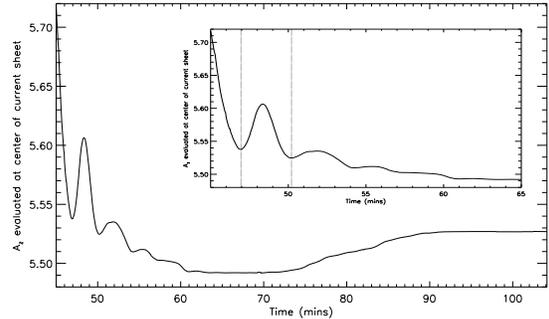}
\caption{Time evolution of $A_z$ located at the center of our (moving) current sheet. Insert shows a blow-up  over a shorter time period. Vertical dotted lines denote how we measure the period of oscillation.}
\label{Figure6}
\end{figure}

\subsection{Quantitative analysis of periodicity}\label{sec:period}

In \S\ref{sec:outflows}, we reported on the formation of strong, quasi-periodic, vertical outflows generated by oscillatory reconnection. As detailed in McLaughlin et al. (\citealp{McLaughlin2009}), by considering the evolution of the vector potential at the center of the current sheet, one  can obtain a quantitative measure of the oscillatory nature of the system. In our numerical experiment, we track the center of the (rising) current sheet and plot the value of the vector potential at that point versus  time. This evolution can be seen Figure \ref{Figure6}.

\begin{figure*}
\epsscale{2.4}
\hspace{-4.0cm}
\plotone{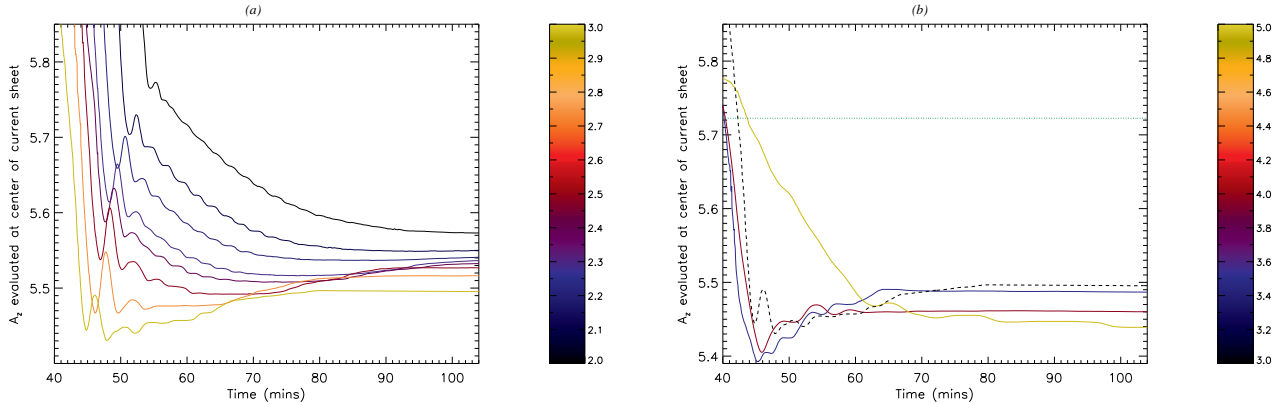}
\hspace{-4.0cm}
\caption{{\emph{(a)}} Parametric study of the evolution of $A_z$ located at the center of the current sheet for eight values ranging from $2\le B_0^* \le 3$, i.e. $2.6\times 10^3 \le B_0 \le 3.9 \times 10^3\:$G. Note that the colour bar indicates the value of the non-dimensional $B_0^*$. {\emph{(b)}} Parametric study of the evolution of $A_z$ for  $B_0^* = 1.9$  (failed emergence, green line),  $B_0^* =  3.5$ (blue line),   $B_0^* =  4$ (purple) and $B_0^* =  5$ (yellow). $B_0^* =  3$ is plotted as a dashed line to aid comparison between both subfigures.}
\label{Figure7}
\end{figure*}

For $45 \lesssim t \lesssim 60\:$mins, the oscillatory nature of our system is clear, and changes in the vector potential can be related to changes in connectivity of the system (see \S$4$ of  McLaughlin et al. \citealp{McLaughlin2009}). Meanwhile, for $60 \lesssim t \lesssim 70\:$mins, the vector potential changes very little and then increases ($70 \lesssim t \lesssim 90\:$mins) to a plateau  ($t \gtrsim 90\:$mins) to a constant $A_z=5.527$. The oscillations between  $45 \lesssim t \lesssim 60\:$mins are directly related to the oscillatory reconnection mechanism at work in our simulations (i.e. a local effect, searching for an equilibrium) whereas the behaviour for  $60 \lesssim t \lesssim 70\:$mins and beyond are related to the  system settling into equilibrium on a global scale. There is no change in the vector potential for $t \gtrsim 90\:$mins.

Since we are primarily interested in oscillatory reconnection in this paper (as opposed to the general behaviour of flux emergence) we shall focus on the time period  $45 \lesssim t \lesssim 60\:$mins, and a blow-up of the evolution of $A_z$ over this time is shown in the insert of Figure \ref{Figure6}. We measure the period of oscillation by considering two local extrema in the vector potential (indicated by the vertical dashed lines in the insert). We measure a period of $195\:$ seconds, i.e.  $3.25\:$mins, for the parameters and initial conditions we have considered.


\section{Parameter Study of $B_0^*$}\label{sec:parameter_study}

We now investigate the dependence of our system to varying $B_0$, i.e. the initial magnetic strength of our buoyant flux tube. This will allow us to see how the period of oscillation changes and investigate how robust our system is to the onset of oscillatory reconnection.

\begin{figure*}
\epsscale{2.4}
\hspace{-4.0cm}
\plotone{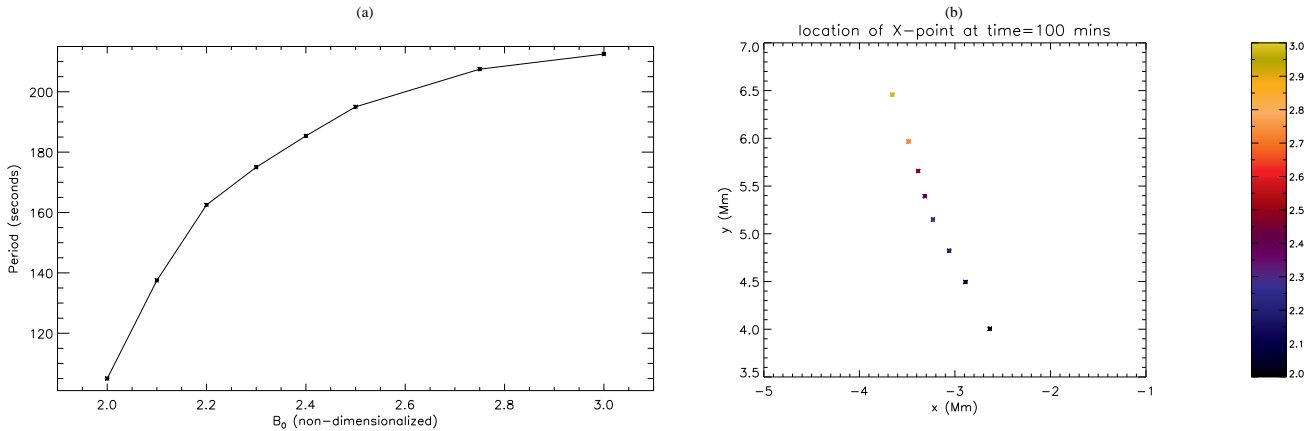}
\hspace{-4.0cm}
\caption{{\emph{(a)}} Parametric study of period of oscillation for  eight values ranging from $2\le B_0^* \le 3$, i.e. non-dimensional $B_0^*$. {\emph{(b)}} Parametric study of the location of the X-point after $t=100\:$mins for eight values ranging from $2\le B_0^* \le 3$. Note that the colour bar indicates the value of the non-dimensional $B_0^*$ and the choice of axes are only a subset of the total numerical domain.}  
\label{Figure8}
\end{figure*}

Figure \ref{Figure7}a shows the full  time evolution of $A_z$ (located at the center of the current sheet) for several different values of the initial value of the magnetic field strength of the buoyant flux tube, i.e. we vary the value of $B_0$. Since this is a numerical parameter study, we choose to present values associated with $B_0^*$, i.e. the non-dimensionalised initial magnetic field strength. Specifically, Figure \ref{Figure7}a reports the evolution of eight values ranging from $2.6\times 10^3 \le B_0 \le 3.9 \times 10^3\:$G or, in non-dimensional units,  $2\le B_0^* \le 3$. Note that {\S\ref{sec:fluxemergence}} above corresponds to  $B_0= 3.25 \times 10^3\:$G, i.e. $B_0^*=2.5$. From the resultant curves, it is clear that oscillatory behaviour is present in all of these numerical experiments. Thus, we can measure the corresponding period of oscillation for each choice of $B_0$ and this is shown in Figure \ref{Figure8}a.

We also investigate larger and smaller values of $B_0$ and these results can be seen in Figure \ref{Figure7}b, which presents values of $B_0 = 2.47$, $4.55$,  $5.2$ and $6.5 \times 10^3\:$G, i.e. $B_0^* = 1.9$, $3.5$, $4$ and $5$. These choices yield behaviors of a significantly different nature to those seen in Figure \ref{Figure7}a. This is for two reasons: firstly, for values of $B_0^*<2$, i.e. $B_0<2.6 \times 10^3\:$G, the initially-submerged flux tube does not successfully emerge into the solar atmosphere. This is in agreement with the results of Murray \& Hood (\citealp{Murray2006}) who found that for low initial magnetic field strengths, the tube cannot fully emerge into the atmosphere since the buoyancy instability criterion is not satisfied. Thus, in this model, for values of $B_0<2.6 \times 10^3\:$G, we have no flux emergence (or failed emergence) and thus no onset of oscillatory reconnection. Thus, high in the solar atmosphere, $A_z$ remains constant. In Figure \ref{Figure7}b, we plot $A_z$ {{corresponding to a $B_0^* = 1.9$  failed emergence (constant value  green line)}} for comparison.

Figure \ref{Figure7}b also shows the time evolution of the vector potentials corresponding to  $B_0^* =  3.5$ (blue line),   $B_0^* =  4$ (purple) and  $B_0^* =  5$ (yellow). We note that the evolution of $A_z$ is again significantly different to that seen in Figure \ref{Figure7}a. $B_0^* =  3$ is plotted as a dashed line to aid comparison between the subfigures. The reason for this significant change in behaviour is that for  $B_0^*>3$, we have successful flux emergence and current sheet/ X-point formation in a similar manner to that detailed in {\S\ref{sec:fluxemergence}}  above but, critically, these strong current sheets now eject {\emph{plasmoids}} from their ends and this fundamentally changes the properties of the current sheet/X-point, since the plasmoids take magnetic flux with them as they leave the current sheet/X-point. Thus, although we still have oscillatory behaviour present in the behaviour of $A_z$ corresponding to $B_0^*=3.5$ and $4$, this represents a fundamentally different regime to that seen for  $2\le B_0^* \le 3$ in Figure \ref{Figure7}a. The effect is even more pronounced for $B_0^*=5$ and beyond.


Let us now investigate the period associated with our choice of $B_0^*$ (Figure \ref{Figure8}a) where we restrict ourselves to  $2\le B_0^* \le 3$ (as explained above, values above and below these limits correspond to fundamentally different regimes). We find that the period of oscillation increases with $B_0^*$, for periods $105-212.5\:$seconds,    i.e. $1.75-3.5\:$min.

Finally, we investigate how the final location of the X-point depends upon our choice of $B_0^*$. We find that the location X-point at $t=100\:$mins, i.e. the final/resting location, is displaced both higher and further to the left for stronger values of $B_0^*$ (again we restrict  ourselves to $2\le B_0^* \le 3$). This behaviour can be seen in Figure \ref{Figure8}b.



\section{Discussion and Conclusions}\label{sec:conclusions}

We have performed numerical experiments of magnetic flux emerging into a coronal hole, modeled as a pre-existing unipolar magnetic field, within a stratified solar atmosphere, and solve the compressible and resistive MHD equations using a Lagrangian remap, shock capturing code: {\emph{LARE2D}}. The long-term evolution of the system is followed and we investigate  how the reconnecting magnetic systems behave as they search for an equilibrium. We find that the initial rise and expansion of the emerging magnetic flux tube are in good agreement with that reported in the existing literature (Figure \ref{Figure2}). Reconnection is initiated at $t=41.6\:$mins {{(for $B_0^*=2.5$ system)}} in which inflows acting perpendicular to the current sheet bring magnetic field into reconnection region, and magnetic flux is ejected at the ends of the current sheet. We find that oscillatory reconnection occurs in the model   (Figure \ref{Figure3}) i.e. a process in which resistive diffusion at the X-point is coupled to global advection of the outer field. We find that the first {\emph{reconnection reversal}}, i.e. change from an orientation 1 current sheet to an orientation 2, occurs around  $t \approx 47\:$mins. The mechanism for oscillatory reconnection is well described by McLaughlin et al. (\citealp{McLaughlin2009}) and Murray et al. (\citealp{Murray2009}) and occurs due to a local imbalance of forces, primarily the gradients in thermal pressure, between the neighbouring flux systems.

We find strong horizontal outflows from both ends of current sheet and, once oscillatory reconnection is initiated, these change direction periodically. Similarly, we find strong vertical outflows in the positive $y-$direction emanating from the upper end of our current sheet, and these sit side-by-side with negative $y-$direction inflows bringing magnetic flux into the reconnection region. {{The direction of these positive/negative velocities changes as the system changes from orientation 1 to  2 (Figure \ref{Figure4}) but we also find on average  there is a vertical upflow directed in the positive  $y-$direction (Figure \ref{Figure5}b).}}

We find that the vertical outflows/collimated jet will be displaced horizontal during  the shortening and lengthening of the evolving currents sheets which results from the oscillatory reconnection mechanism. This displacement gives the vertical outflow jets a swaying nature and explains the transverse nature of the oscillations. This was further confirmed by analysing  ${{{{v}}}}_x(-2.5\:{\rm{Mm}}, 7\:{\rm{Mm}},t)$, which clearly demonstrated  the transverse behaviour (Figure \ref{Figure5}a).

In order to quantitatively estimate the vertical outflows, we measured the vertical velocity at a fixed point: ${{{{v}}}}_y(-2.5\:{\rm{Mm}}, 7\:{\rm{Mm}},t)$. Here, the positive (material ejected from the end of the current sheet) and negative (inflow into the reconnection region) vertical flows were visible and as well as the periodic nature of the oscillations (Figure \ref{Figure5}b). Only a few complete periods were observed   and so we labelled this quasi-periodic behaviour. We fit the damped oscillations with exponential envelopes and found that there was a preference for positive velocity over negative, i.e. evidence of a  positively-directed vertical net flow in the model. {{Note that the generation of this  positively-directed vertical net flow is tightly localised to the upper ends of the (evolving) current sheet and so similar results are obtained only for  other choices of fixed points located around   $(-2.5\:$Mm, $7\:$Mm) and within the same domain of connectivity.}}



By  tracking the vector potential, $A_z$, at the center of the (moving) current sheet, we quantitatively measure the period of oscillation, allowing us to measure a period of $195\:$seconds  ($3.25\:$mins) for an initial magnetic flux tube strength of $B_0=3.25 \times 10^3\:$G ($B_0^*=2.5$). 


We also perform a parameter study varying the initial magnetic strength of the buoyant flux tube, i.e. $B_0$. We find that for a range of parameters applicable to the solar atmosphere, $2.6\times 10^3  \le B_0 \le 3.9 \times 10^3\:$G, we observed similar flux emergence and oscillatory reconnection behaviour, with each $B_0$ corresponding   to its own period of oscillation in the range $1.75 \le {\rm{period}} \le 3.5\:$mins. Essentially, the stronger the initial flux tube strength, the longer the period of oscillation.


However, for $B_0 \le 2.47 \times 10^3\:$G, we do not observe successful flux emergence and thus, obviously, there is no subsequent  oscillatory motion. This is in agreement with Murray \& Hood (\citealp{Murray2006}) who found that for low initial magnetic field strengths, the tube cannot fully emerge into the atmosphere since the buoyancy instability criterion is not satisfied, i.e. \lq{failed}\rq{ } flux emergence. Further details of the buoyancy instability criterion can be found in Newcomb (\citealp{Newcomb1961}), Yu (\citealp{Yu1965}), Thomas \& Nye (\citealp{Thomas1975}), Acheson (\citealp{Acheson1979}) and Arcontis et al. (\citealp{Vasilis2004}). Thus, it is the buoyancy instability criterion that dictates the lower limit in our model, given our choices of parameters. For different parameters, e.g. a stronger/weaker equilibrium uni-directional magnetic field, the buoyancy instability criterion will have a higher/lower threshold, although a full investigation must be  undertaken to investigate the true behaviour.

We also investigate larger values, i.e. $B_0 \ge 4.55 \times 10^3\:$G. Here, unlike for $2.6\times 10^3  \le B_0 \le 3.9 \times 10^3\:$G, we observe the formation of plasmoids ejecting from the ends of our current sheet. These ejected plasmoids change the properties of the X-point, e.g. taking magnetic flux with them. Thus, even though we still have oscillatory behaviour, seen in the evolution of $A_z$, this represents a fundamentally different regime than that of  $2.6\times 10^3  \le B_0 \le 3.9 \times 10^3\:$G. The exact reason for plasmoid formation above a particular threshold strength is uncertain, and will be investigated in future work.

Finally, we investigated how the final location of the X-point depends upon our choice of $B_0$. We found that the location X-point at $t=100\:$mins, i.e. the final/resting location, was displaced both higher and further to the left for stronger values of $B_0$ (we restricted our records to  $2.6\times 10^3  \le B_0 \le 3.9 \times 10^3\:$G). Both the longer duration periods of oscillation and increased displacement/height of the X-point can be fully explained since stronger $B_0$ tubes have larger emergence velocities and thus greater momenta. Thus, the larger momentum flux tubes are able to compress the X-point to a greater extent (resulting in a stronger/longer current sheer and thus longer periods for the subsequent oscillatory reconnection, which explains Figure \ref{Figure8}a) and, secondly, higher $B_0$ flux tubes with larger momenta carry the tube higher and farther into the atmosphere (explaining Figure \ref{Figure8}b).


Thus, we have presented numerical simulations that {\emph{naturally generate quasi-periodic flows  with a characteristic transverse/swaying aspect}}. Such outputs result from the oscillatory reconnection physical mechanism, i.e. in a self-consistent manner since {\emph{no periodic driver is imposed on our system}}. The vertical speeds of the outflows, $20-60\:$km/s, are comparable to those reported in recent observations. By varying the initial strength of our submerged flux tube, we recover periodicities in the range of $1.75 -  3.5\:$mins.    Thus, the   mechanism of oscillatory reconnection may provide a  physical explanation for the generation of some of the recent quasi-periodic, transverse, vertical  motions/jets/outflows reported by a multitude of authors (see \S\ref{sec:Introduction} for details). In particular, the oscillatory reconnection mechanism presented here may explain the observations and simulations by  Nishizuka et al. (\citealp{Nishizuka2008}; \citealp{Nishizuka2011}) and  He et al. (\citealp{He2009a}; \citealp{He2009b}), who detail {\emph{Hinode/SOT}} observations and simulations of transverse motions on a spicule, originating from the cusp of an inverted Y-shaped structure, as well as more recent work by Ding et al. (\citealp{Ding2011}) and Harra et al. (\citealp{Harra2011}).

The oscillatory mechanism presented here may also partially explain quasi-periodic pulsations  (see reviews by Aschwanden \citealp{Aschwanden2003}; Nakariakov \& Melnikov \citealp{Nakariakov2009}). Such oscillatory behavior has been observed in a number of solar and stellar flares (Mathioudakis et al. \citealp{Mathioudakis2003}; \citealp{Mathioudakis2006}; McAteer et al. \citealp{McAteer2005}; Inglis et al. \citealp{Inglis2008}; \citealp{Inglis2009}; Nakariakov et al. \citealp{Nakariakov2010};  \citealp{Nakariakov2011}) but the true physical mechanism responsible remains uncertain.

Finally, it should also be noted that the mechanism generates both ${{{{v}}}}_x$ and ${{{{v}}}}_y$  motions together (one cannot exist without the other), and such motions are exponentially damped. Thus, if such signals are detected, they may be decaying not due to a damping mechanism but {\emph{due to the generation mechanism itself}}. This is not surprising given that flux emergence injects a finite amount of energy into the (oscillatory reconnection) mechanism, and so it is expected that     the phenomena and outputs  will also only be of a finite duration, i.e. this is a {\emph{dynamic}} reconnection phenomenon as opposed to the classical steady-state (time-independent) reconnection models.

Although we have only presented  a specific example of reconnection initiated by flux emergence, we believe the oscillatory reconnection mechanism described in this paper is a robust, general phenomenon that may be observed in other systems that demonstrate finite-duration reconnection. Further studies should focus on the inclusion of heat conduction (e.g. Miyagoshi \& Yokoyama \citealp{Miyagoshi2003}; \citealp{Miyagoshi2004}) which is expected to reduce the temperature of the outflow jets. However, the density of outflows will also be increased by heat conduction, i.e. to ensure force balance in the current sheet, which may make the outflow jet more observable (e.g. Shiota et al. \citealp{Shiota2005}). Finally, the oscillatory reconnection mechanism itself should be investigated further  (e.g. Gruszecki et al. \citealp{Marcin2011}) and extended to fully three-dimensional studies. Evidence of oscillatory reconnection in 3D flux emergence simulations was recently reported by Archontis et al. (\citealp{VasilisOSCILLATORYRECONNECTION}).

\acknowledgments

\section*{Acknowledgments}

JM  wishes to thank M Murray, V Archontis and A Hood for helpful and insightful discussions and suggestions.   RE acknowledges M. K\'eray for patient encouragement and is also grateful to STFC (UK) and NSF, Hungary (OTKA, Ref. No. K83133).  The authors  also acknowledge IDL support provided by STFC. The computational work for this paper was carried out on the joint STFC and SFC (SRIF) funded cluster at the University of St Andrews (Scotland, UK).

\end{document}